\documentclass{article}
\usepackage[style=chem-acs, backend = biber, ]{biblatex}
\usepackage{lipsum,capt-of,graphicx}
\usepackage{hyperref}
\usepackage{comment}
\usepackage[a4paper, total={6in, 8in}]{geometry}
\usepackage[version=4,arrows=pgf-filled,textfontname=sffamily,mathfontname=mathsf]{mhchem}
\usepackage{bookmark}
\title{Quantifying non-Gaussian diffusion in transient microscopy using excess kurtosis.}

\author{Enrique Arévalo Rodríguez$^1$$^2$, Marc Meléndez Schofield$^1$$^2$, Jorge Cuadra$^1$$^2$,
\\ Ferry Prins$^1$$^2$}
\date{%
$^1$ Department of Condensed Matter Physics, Autonomous University of Madrid, Madrid, Spain\\%
$^2$ Condensed Matter Physics Center, IFIMAC, Madrid, Spain\\[2ex]%
}

\begin{document}
\maketitle

\begin{abstract}
Research on energy transport has advanced in recent years with the emergence of transient microscopy techniques that allow for imaging of carriers with high spatial and temporal resolution. These techniques often rely on Gaussian fits to quantify the broadening of the population, however, this can lead to the misrepresentation of results when multiple, overlapping species coexist. Transient scattering microscopy (TScM), has emerged as an alternative to traditional techniques. However, the sensitivity of TScM to different carriers accentuates the limitations of traditional Gaussian fits. In this work we use TScM to visualize exciton transport in bulk TMDCs and show that exciton populations exhibit non-Gaussian profiles by analyzing their excess kurtosis. Numerical simulations incorporating anomalous diffusion -such as Meitner-Auger recombination and trap states- reproduce these experimental observations. Furthermore, by tuning the injected carrier density, we demonstrate that the temporal signature of the kurtosis is distinct for Auger-dominated and trap-dominated regimes. Additionally, we find that traditional Gaussian-fitting methods can yield inconsistent results for the extracted diffusivities. As an alternative, we implement a discrete variable calculation which yields robust, consistent diffusivity values. Our results establish kurtosis as a vital diagnostic parameter for identifying anomalous diffusion and demonstrate the necessity of moving beyond Gaussian approximations for accurate analysis of TScM data.

\end{abstract}

\section{Introduction}

The study of carrier transport has advanced significantly in recent years with the emergence of a series of transient microscopy techniques, which enable imaging of energy carriers with nanometer-scale spatial and sub-nanosecond temporal resolution\cite{akselrod_subdiffusive_2014}\cite{ginsberg_spatially_2020}\cite{shi_exciton_2013}\cite{afrin_rapid_2024}\cite{saidaminov_multi-cation_2020}\cite{nagaya_wong_robust_2022}. These time-resolved measurements offer crucial insights into the various transport regimes that carriers may undergo during their lifetime, ranging from ballistic to diffusive and subdiffusive behaviors\cite{tulyagankhodjaev_room-temperature_2023}\cite{seitz_exciton_2020}\cite{seitz_mapping_2021}. By providing spatiotemporal maps of carrier dynamics, detailed information on how carriers evolve across complex energy landscapes can be obtained. Despite their relatively recent introduction, transient microscopy techniques have rapidly become an essential tool for the optoelectronic characterization of semiconductors.

Different transient microscopy techniques distinguish themselves mainly through their respective contrast mechanisms. Early examples include transient photoluminescence\cite{irkhin_direct_2011} and transient absorption microscopy\cite{ruzicka_exciton_2012}, both of which use diffraction limited excitation of a carrier population followed by spatiotemporal tracking of this population. In the case of transient photoluminescence microscopy (TPLM), radiative decay from the carriers is detected using spatially resolved time-correlated single photon counting or using streak camera approaches\cite{ziegler_mobile_2023}. Transient Absorption Microscopy (TAM) is a pump-probe technique that uses a spatially resolved probe beam to detect changes in the absorptivity of the material in the presence of carriers. Using absorption as a contract mechanism has a major advantage in its ability to detect carriers independent of their radiative decay efficiencies\cite{deng_long-range_2020}\cite{yuan_twist-angle-dependent_2020}\cite{cui_transient_2014}. However, transient absorption microscopy also has typically inferior signal-to-noise ratios, requiring high excitation powers to obtain meaningful data. 

More recently, transient microscopy based on interferometric scattering was reported\cite{delor_imaging_2020}\cite{delor_carrier_2020}. Transient Scattering Microscopy (TScM, also sometimes referred to as stroboSCAT) bases its contrast on small changes in the refractive index of a material in the presence of carriers. Analogous to traditional interferometric scattering (iSCAT)\cite{ortega_arroyo_interferometric_2016}, TScM benefits from the interference of the scattered probe with reflection at the glass-substrate interface, allowing for higher signal-to-noise ratios as compared to absorption based techniques. Several studies on different material systems have since then been reported, including metal halide perovskites \cite{delor_imaging_2020}, super atomic semiconductors \cite{tulyagankhodjaev_room-temperature_2023} and few layer transition metal dichalcogenides (TMDCs) \cite{zhu_direct_2017}\cite{su_dark-exciton_2022}.  

Importantly though, sensitivity to a large variety of populations can complicate the interpretation of the results. Extracting quantitative information from transient microscopy relies on the extraction of the change in variance of the excited state population, given by $\Delta\sigma^2 = \sigma(t)^2 -\sigma(0)^2$.  To determine the variance, most Transient Microscopy studies fit the different time-slices to a Gaussian function\cite{shcherbakov-wu_persistent_2024}\cite{deng_imaging_2020}\cite{feldman_inverted_nodate}\cite{thiebes_quantifying_2024}. However, as a number studies have shown, complex dynamics can emerge when multiple mobile species are present\cite{kulig_exciton_2018}\cite{seitz_mapping_2021}\cite{magdaleno_role_2025}, in the presence of higher order recombination, or when subpopulations of trapped and free carriers dynamically interchange with each other. Importantly, in TScM, depending on the type of population and its corresponding effect in the refractive index of the material, both positive and negative contrast can be obtained, in some cases both effects can even coexist\cite{weaver_detecting_2023}\cite{utterback_nanoscale_2021}. Careful evaluation of the spatiotemporal evolution in the presence of the coexistence of multiple populations is therefore critical and requires more accurate quantitative analyses that go beyond Gaussian approximations of the carrier distribution.

Here, we employ Transient Scattering Microscopy (TScM) to investigate exciton transport in bulk tungsten diselenide (\ce{WSe2}). Our data reveals significant deviations from pure Gaussian diffusion, which we quantify by means of the kurtosis of the exciton population. The temporal evolution of the excess kurtosis shows a transition from heavy tailed distributions at the beginning to short tailed at the end. Repetition rate dependent measurements show that the former is the result of a surviving population from previous laser pulses, while numerical simulations indicate that the late time dynamics is best described by excitons that encounter shallow traps as they diffuse through the material. Furthermore, by means of power dependent measurements we demonstrate that Meitner-Auger recombination reduces excess kurtosis in the early times. Finally, as traditional Gaussian analysis fail to accurately describe the population evolution, we introduce an alternative and more reliable method to directly extract the spatial variance (and thus the diffusivity) from the distribution. Our work establishes kurtosis as a reliable metric to evaluate the signatures of different transport dynamics before making assumptions about the spatial distribution and its temporal evolution. 

\begin{figure} [ht]
  \includegraphics[width=\linewidth]{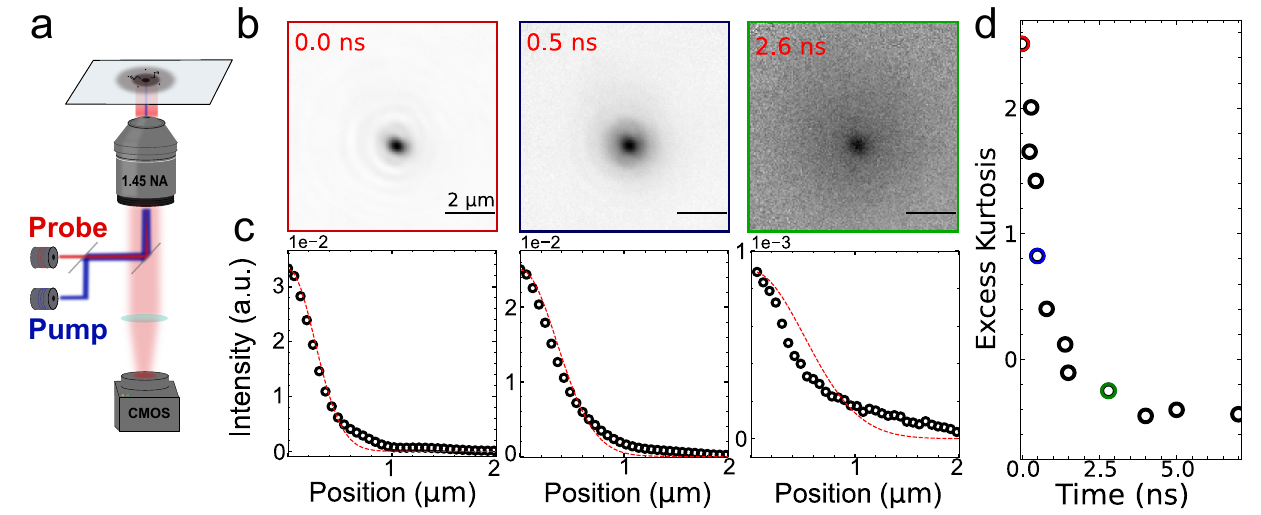}
  \caption{a. Schematic of the transient scattering microscopy setup. b. Transient scattering differential images taken at different pump-probe time delays. c. Corresponding azimuthally averaged profiles of the pictures shown in b., dark points are the resulting averaged data points while dashed red lines show the best gaussian fit for the provided data. d. Excess kurtosis as a function of time for the complete dataset shown in b., kurtosis is calculated using the full 2D image for each time delay to increase SNR.}
\label{fig1}
\end{figure}

\section{Results and discussion}

We perform transient scattering microscopy (TScM) measurements on multilayer tungsten diselenide (\ce{WSe2}) flakes ($>$ 10 layers), exfoliated from commercially obtained large crystals of \ce{WSe2} and transferred to glass cover slips (see experimental methods section for details). TScM is performed using a near-diffraction-limited excitation laser (405 nm) as pump, whereas the probe (780 nm) is focused in the back focal plane of the objective to obtain wide field illumination. Images of the reflected probe are collected using a CMOS camera (FLIR BFS-U3-28S5M-C), a schematic of the optical setup in shown in Fig\ref{fig1}a. Synchronization between the CMOS and the laser driver with electronic delay (Picoquant Sepia PDL 828) allows us to record images of the excitation populations at different pump-probe delay times. Consecutive pump ON and pump OFF images are acquired and subsequently divided to generate differential images in the form $diff = ON/OFF -1$. Figure \ref{fig1}b shows a series of resulting differential images for different pump-probe delays at an excitation fluence of $ 16 \mathrm{\mu J/cm^{2}}$ and 5 MHz laser repetition rate, with each image normalized to its maximum (negative) signal intensity. The negative intensity of the signal indicates a reduction in the reflectivity of the material upon photoexcitation, consistent with a reduced refractive index upon depletion of the ground-state carriers (\cite{ginsberg_spatially_2020}). 

The differential images reveal two contrast sources with distinct spatiotemporal dynamics: a fast-moving population that exhibits clear spatial broadening over time, and a slow-moving, long-lived population that remains largely stationary. As such, the time-dependent spatial distribution of the exciton population exhibits significant deviations from a two-dimensional Gaussian profile, deviating from purely Gaussian diffusion, complicating the quantitative analysis of the exciton diffusion dynamics. Figure \ref{fig1}c presents azimuthal averaged profiles of the exciton density at selected time delays shown in Figure \ref{fig1}b, along with their respective optimal Gaussian fits (dashed lines). At early times, the Gaussian fit fails to capture the heavy tails of the distribution. In contrast, as time goes by, the heavy tails of the distribution gradually vanish, at 0.3 ns the Gaussian fit does a good job of at fitting the profile, but at later times the tails become shorter than expected, leading to a short tailed or flat-topped distribution. We can describe the deviation from Gaussian shapes by calculating the kurtosis of the distribution as a function of time (see Methods for details). Within this definition, a perfect Gaussian distribution has an excess kurtosis of 0, while positive and negative values represent heavy tailed and flat-topped distributions, respectively. Results of the excess kurtosis for each differential image show a consistent temporal evolution, where the excess kurtosis starts positive and becomes smaller as time passes, stabilizing after reaching a negative value of around $-0.5$, as shown in Fig \ref{fig1}d.

\begin{figure}[ht]
  \includegraphics[width=\linewidth]{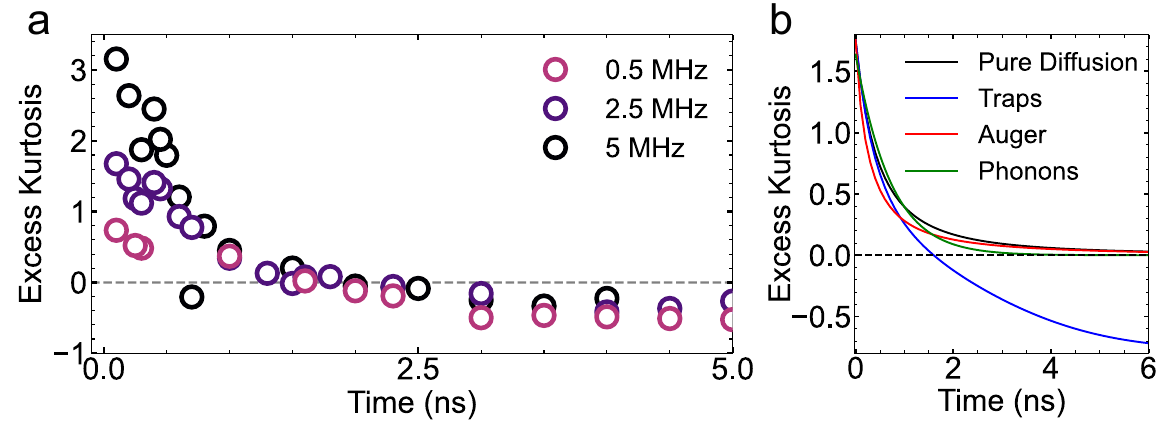}
\caption{ a. Excess kurtosis as a function of time for selected datasets acquired with different laser repetition rates. b. Numerical simulations for different dynamics representing the corresponding evolution of the excess kurtosis as a function of time. Black corresponds to pure Gaussian diffusion with no other effects, the red line represents the dynamics when Meitner-Auger recombination is present, the blue line shows the evolution of the kurtosis for dynamics dominated by trap states and the green line the calculation for a combination of non-interacting heat and exciton populations.}
\label{fig2}%
\end{figure}

To better understand the origin of the early excess kurtosis, we performed TScM at different laser repetition rates varying from 500kHz to 10MHz while maintaining the same fluence ($\mathrm{65 \mu J/cm^2}$). Figure \ref{fig2}a shows the temporal evolution of the excess kurtosis for selected repetition rates. We observe that the early excess kurtosis drops significantly with decreasing repetition rate, going from values $>2$ for 10MHz,  indicating a heavy tailed distribution, to an excess kurtosis $<1$ for 500kHz, indicating a much more Gaussian-like distribution. The decreasing excess kurtosis with decreasing repetition rate suggests that the source of the early positive kurtosis lies in a surviving population from previous laser pulses. This residual contrast appears as a secondary population that merges with the newly generated excitons by the subsequent pump pulse. This effect vanishes when the time between pulses is sufficiently long, allowing long-lived states to relax. This hypothesis is supported by calculations (SI section 1), which demonstrate that the convolution of two Gaussian populations can produce a combined distribution with positive excess kurtosis. 

Having established the origin of the positive kurtosis at early times, we now turn to the spatiotemporal dynamics at longer times. All the results presented thus far exhibit a consistent trend where the excess kurtosis tends to negative values, stabilizing around -0.5. This behavior contrasts with the expectation for Gaussian diffusion. If we consider pure Brownian motion, any initial population, independent of the initial shape of the distribution, will tend to a value of 0 excess kurtosis with time. This behavior is confirmed by numerical simulations, as shown in Figure \ref{figS2}. Consequently, the time-dependent evolution towards a negative excess kurtosis at long times is a direct indication of more complex dynamics than simple Gaussian diffusion through Brownian motion.


Negative excess kurtosis is indicative of a flat-top distribution and has in the past been associated with Meitner-Auger recombination.\cite{wietek_nonlinear_2024}\cite{perea-causin_exciton_2019} As Meitner-Auger recombination occurs more efficiently at high exciton densities, the resulting faster decay at the center of the population indeed leads to a flattening of the distribution. Crucially though, the effect of Meitner-Auger recombination is limited to early times, diminishing quickly as the exciton density reduces as a result of both population decay and diffusion processes. As time goes by, Gaussian diffusion would again tend to a value of 0 excess kurtosis, as confirmed by numerical simulations (see red solid line in Figure \ref{fig2} b).  

We now turn to the possibility of multiple mobile species causing negative excess kurtosis. As mentioned, TScM is sensitive not just to excitons or free carriers, but also to phonon populations. Phonons are generated as photoexcited carriers undergo bandedge relaxation in the first ps after excitation. The spatiotemporal evolution of the contrast is then a convolution of both fast-moving short-lived excitons and slow-moving long-lived phonons. As time goes by, this would lead to a short-lived heavy tailed distribution caused by the fast outward moving excitons, leaving the slow moving phonons behind. With time, however, the long-lived phonon population with normal Gaussian diffusion would take over, once more leading to a negligible excess kurtosis. Simulations of this scenario are represented by the green solid line in Figure \ref{fig2} b). Indeed, at early times, a small prolongation of the positive excess kurtosis is observed, decaying to 0 as time goes by. 

A distinct scenario with multiple mobile species is the presence of shallow traps. The presence of trap states has been shown to significantly affect the spatiotemporal dynamics of the population in transient microscopy measurements \cite{seitz_mapping_2021}\cite{akselrod_subdiffusive_2014}\cite{lo_gerfo_morganti_transient_2025}. In contrast to co-propagating populations of excitons and phonons, shallow traps lead to a dynamic interchange between free and trapped excitons through consecutive trapping and detrapping events. Importantly, free excitons have a relative fast decay compared to trapped excitons. The less time excitons have resided in traps,  the faster they move and the shorter their lifespans. This scenario indeed leads to a short-tailed (flat-topped) distribution with negative excess kurtosis, as demonstrated by our simulations (solid blue line in Figure \ref{fig2} b). 

From our simulations, we can therefore conclude that the negative excess kurtosis is likely the result of the presence of shallow traps. The long-lived population responsible for early positive kurtosis could have the same origin, as traps cause longer-lived populations that can survive until the next laser pulse arrives \cite{magdaleno_role_2025}. Importantly though, long-lived phonon populations provide an alternative explanation for the early positive kurtosis. Our analysis highlights the challenge that sensitivity to different carriers brings and the importance of kurtosis as an indicator of non-Gaussian diffusion dynamics.

\begin{figure}[ht]
\includegraphics[width=0.45\linewidth]{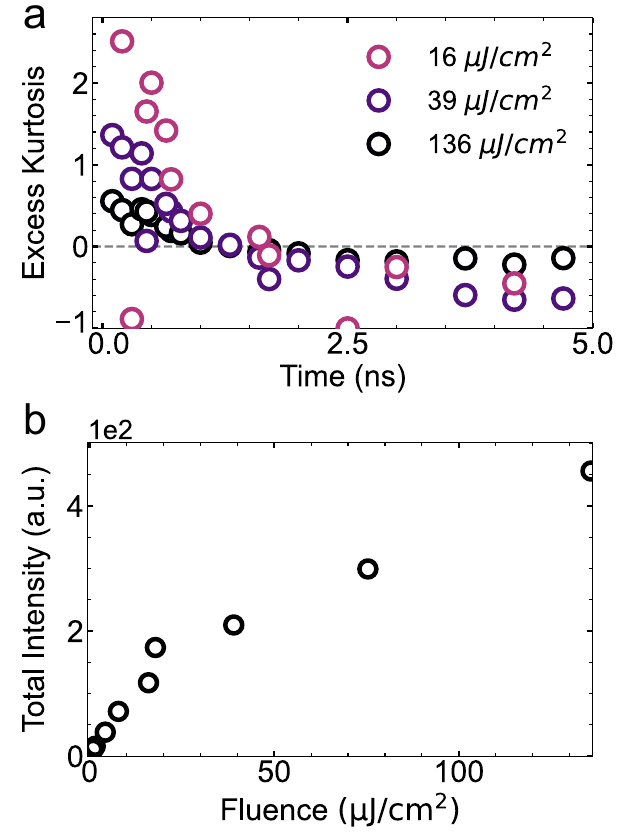}
\caption{a. Excess kurtosis as a function of time for different injected carrier densities. b. Total calculated intensity contrast for the measured laser fluences.}
\label{fig3}
\end{figure}

While it is clear that long-lived states are the dominant factor in determining the evolution of excess kurtosis, Meitner-Auger recombination is known to play an important role in TMDCs at higher excitation densities \cite{kulig_exciton_2018}\cite{perea-causin_exciton_2019}. To elucidate how Meitner-Auger recombination additionally affects the excess kurtosis in \ce{WSe2}, we perform TScM for increasing laser fluences, varying from 16 $\mathrm{\mu J/cm^2}$ to 136 $\mathrm{\mu J/cm^2}$. These experiments were performed at a repetition rate of 5 MHz. As can be seen in Figure \ref{fig3}a, the increasing excitation density has a distinct effect on the kurtosis, reducing the early time excess kurtosis as fluence increases. This trend is qualitatively consistent with our numerical simulations (Figure \ref{fig2}b), which also show a rapid decrease in early-time kurtosis. To corroborate that this trend originates from Meitner-Auger recombination, we calculated the total contrast intensity by integrating the signal across all pump-probe delays for each fluence. The results (Figure\ref{fig3}b) reveal a clear transition from linear growth at low fluences ($<$10 $\mathrm{\mu J/cm^{2}}$) to a sublinear regime at higher fluences, indicating a saturating behavior characteristic of Auger-mediated exciton decay.

\begin{figure}[ht]
\includegraphics[width=0.45\linewidth]{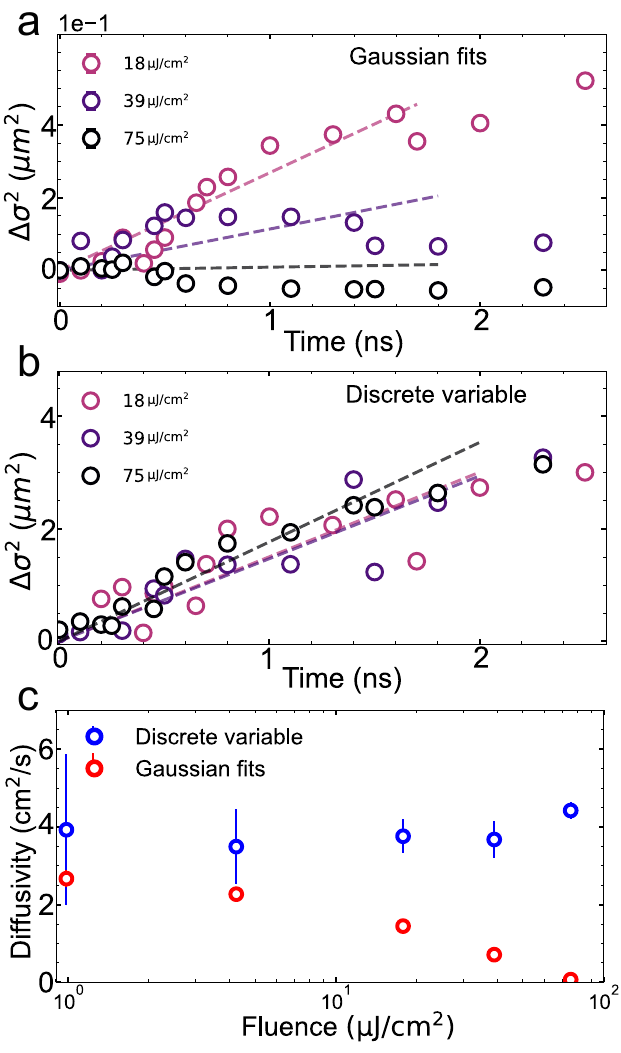}
\caption{ a. Extracted values for $\Delta \sigma^2$ using one-dimensional gaussian fits from linecuts for selected injected energy densities. b. Variances extracted from the same datasets as a. using the discrete variable approach on the two-dimensional distributions. c. Comparison of the calculated diffusivities for each method at different incident fluences, all diffusivities are obtained by performing a linear fit on the first 1.5 ns, before the sublinear diffusion regime.}
\label{fig4}
\end{figure}

A critical question is how strongly an analysis based on conventional Gaussian fits is affected by distributions with non-zero excess kurtosis. As it is common in the field \cite{ziegler_excitons_2022}\cite{yuan_twist-angle-dependent_2020}\cite{baxter_coexistence_2023}\cite{shi_direct_2022}, we perform Gaussian fits to one-dimensional line cuts from the differential images to calculate the change in variance as a function of time ($\mathrm{\Delta \sigma^2 = \sigma^2(t) - \sigma^2(t=0) }$). As shown in Figure \ref{fig4}a, using Gaussian fits yields an unexpected dependence on the excitation power, with higher powers almost completely eliminating spatial broadening of the population. This result suggests clear limitations of Gaussian fitting when excess kurtosis is strongly time-dependent. 

To avoid making any assumptions on the shape of the carrier distribution, we can calculate the $\mathrm{\sigma^2}$ of the distribution by calculating the variance using a discrete variable approach:

\begin{equation}
    \sigma^2 = \frac{1}{I} \sum^n_if_i(x_i-\mu)^2
    \label{eq_vars}
\end{equation}

where $I$ is the total intensity (sum of all the pixel intensities), $x_i$ corresponds to the position of each pixel, $f_i$ is the intensity value in that pixel and $\mu$ the expected value for x. In contrast to the Gaussian method, the discrete variable approach consistently recovers a comparable evolution for $\Delta \sigma^2$ across all excitation powers (Figure \ref{fig4}b). Figure \ref{fig4}c summarizes the differences in the extracted diffusivity for both methods (see linear fits in Figure \ref{fig4}a and b and methods section for details). While Gaussian fits yield progressively smaller diffusivities at higher powers, the discrete variable method shows diffusivity values stable around 4 $\mathrm{cm^2/s}$. This value is comparable to previous reports on TMDCs \cite{goodman_substrate-dependent_2020}\cite{cordovilla_leon_exciton_2018}. Our results validate the discrete variable approach as a more robust and consistent alternative to the more conventional Gaussian fitting.

\section{Conclusion}

We have presented the direct visualization of the exciton transport dynamics in bulk \ce{WSe2} using transient scattering microscopy. Our results show that the time-dependent exciton distributions deviate significantly from Gaussian diffusion and we quantify this deviation using kurtosis. Using repetition rate and fluence dependence of the excess kurtosis, we identify how trap states and Meitner-Auger recombination influence the observed population distributions at different points in time.  We show how Gaussian fits can produce unphysical results for the diffusivity in the case of excess kurtosis and, alternatively, we propose the use of a discrete variable approach to avoid such artifacts. Our results establish the temporal evolution of the kurtosis as a sensitive diagnostic tool that can serve as an important first step in the analysis of Transient Microscopy data to identify non-Gaussian diffusion and avoid misrepresentation of the diffusivity of excited state carriers.

\section{Experimental Section}
\subsubsection{Sample preparation}

Many-layer \ce{WSe2} flakes were exfoliated from a large crystal following reported techniques \cite{castellanos-gomez_deterministic_2014}. Briefly, after exfoliation using adhesive tape, the flakes were deposited onto a transparent polymethylmethacrylate (PMMA) stamp. A three-axis manipulator was then used to transfer the \ce{WSe2} onto the desired glass substrate. The thickness of the flakes is estimated to be $\sim20$ layers by optical inspection.

\subsubsection{Transient scattering microscopy}

For transient scattering measurements, laser diodes were used for the pump (405 $nm$ Picoquant) and probe (780 $nm$ Picoquant) sources, both lasers are controlled via the same driver (Picoquant Sepia II) which allows control of the delay between the two lasers via its electronic capabilities. After spatially filtering both lasers the two are combined using a dichroic mirror and sent to the objective (Nikon Plan Apo 1.45 NA 100x). The reflected light is sent back to a CMOS camera after filtering out remaining pump light. The pump laser is modulated at 440 Hz, while the CMOS acquires pictures at twice the speed (880Hz), thus acquiring consecutive images with and without the pump excitation.

We analyze our images by dividing subsequent ON/OFF image pairs, thus obtaining differential images $Diff = ON/OFF-1$, which contain information about how much the reflection of the material changes between ON and OFF cases. A single experiment typically involves 4000 pairs for each pump-probe delay, the pairs are averaged together to reduce noise as much as possible. 

\subsubsection{Derivation of the kurtosis}

Kurtosis is conventionally defined in one dimension as 
\begin{equation}
    k = E[\frac{(x - \mu_x)^2}{\sigma^4}]
    \label{eq_kurt_1d}
\end{equation}

Although the kurtosis is well defined in one dimension, for a bivariate distributions there are several ways to define the kurtosis. For isotropic distributions, where the variance in both directions is equal ($\sigma^2_x = \sigma^2_y$), we can define the kurtosis in the form \ref{eq2}

\begin{equation}
    k = E[( \frac{ ( (x-\mu_x)^2 + (y- \mu_y)^2 )^2 }{\sigma^4})]
    \label{eq2}
\end{equation}

where E is the distribution function and $\mu_x$ and $\mu_y$ are the expected values for the x and y positions, the value of $\sigma^2$ is calculated using the discrete variable approach. It is worth noting that for this definition the kurtosis for a Gaussian distribution is 2 instead of the typical 3 for the one dimensional case. With the exception of the data presented in Figure S9, where kurtosis is calculated one-dimensionally for azimuthally averaged profiles and $EK = k - 3$, all excess kurtosis values are calculated from the two-dimensional distributions as $EK = k - 2$. \cite{mardia_measures_1970}


\medskip

\textbf{Acknowledgements} \par 
This work was funded by the European Union (ERC, EnVision, project number 101125962). Views and opinions expressed are however, those of the author(s) only and do not necessarily reflect those of the European Union or the European Research Council Executive Agency. Neither the European Union nor the granting authority can be held responsible for them. We acknowledge additional funding from the Spanish AEI under grant agreements PID2022-141579OB-I00, TED2021-131018B-C21, and CNS2023-143577. In addition, we acknowledge the support from the “(MAD2D-CM)-UAM” project funded by Comunidad de Madrid, by the Recovery, Transformation and Resilience Plan, and by NextGenerationEU from the European Union. Finally, we acknowledge support from the “María de Maeztu” Program for Units of Excellence in R\&D (CEX2023-001316-M). 

We thank D. Xu and M. Delor as well as J.K. Utterback for discussions and M. Frising for technical assistance.

\medskip
\printbibliography
\medskip

\section{Supporting Information} \par 
\subsection{Numerical simulations.}

In the main text. We performed numerical simulations of exciton and heat diffusion. These simulations numerically integrate the following coupled differential equations. which determine how free, $c_0(r,t)$, and trapped, $c_i(r,t)$ excitons propagate:

\begin{equation}
    \frac{\partial c_0(r,t)}{\partial t} = D_0\nabla^2c_0 - (\nu_0 + \sum^N_{i=i} \lambda_i) c_0 + \sum^N_{i=i} \mu_ic_i + \rho c_0^2
\end{equation}

\begin{equation}
    \frac{\partial c_i(r,t)}{\partial t} = \lambda_ic_0 - (\mu_i + \nu_i)c_i
    \label{traps_eq}
\end{equation}

In these equations, $D_0$ is the exciton diffusivity, $\nu_0$ the free exciton recombination rate, $\lambda_i$ the trapping rate into trapped state $i$, $\mu_i$ the corresponding detrapping rate, $\nu_i$ the trapped state recombination rate and $\rho$ the Meitner-Auger recombination coefficient. The main text shows results for simulations in four different scenarios; pure diffusion, trap states, Auger and phonons, for each of these a different combination of the active terms is used.

For simulations (Figure\ref{fig2}b) incorporating only pure diffusion these are the only terms taken into account, and the positive excess kurtosis is fixed by changing the shape of the starting distribution to better match the experimental observation.

In the case of Auger recombination, we incorporate exclusively the non linear term ($\rho c_0^2$). Similarly to before, the starting excess kurtosis is manually adjusted to fit the experiments.

For the simulations incorporating trap states, we introduce trapping ($-\lambda_1c_0$) and detrapping ($+\mu_1c_1$) events, thus incorporating the second coupled differential equation (eq.\ref{traps_eq}). Again, the excess kurtosis at the beginning is manually adjusted.

Lastly, for phonon transport, the model maintained the pure diffusion equation but we introduced a second non-interactive population with a lower diffusivity and longer lifetimes. These are chosen to reproduce experimental results. The total population calculated as the sum of both of these, independent, distributions. Crucially, in this case, instead of adjusting the positive excess kurtosis at early times we tune the exciton-phonon ratio to match it.

Importantly, all the simulations generate one-dimensional profiles as such, to calculate the variances and kurtosis from the simulated distributions we use equations \ref{eq_vars} and \ref{eq_kurt_1d} respectively.

\subsection{Excess kurtosis for the sum of two Gaussian profiles.}

In Figure\ref{fig2} of the main text, we show that positive values of the early excess kurtosis may be caused by the overlap of two coexisting populations, a main population generated by the excitation beam and a surviving population originating from previous laser pulses. It is possible to mathematically prove that the combination of two populations will lead to positive values of the excess kurtosis. 

Let us assume a total population described as $p(x) = a p_1(x) + b p_2(x)$, where $p_1$ and $p_2$ are the two coexisting populations, $a,b \ge 0$ and $a+b =1$. To find the kurtosis, we first need to calculate the variance $\sigma^2$.
\begin{equation}
    \sigma^2 = \int^ \infty_\infty p(x)x^2dx = a\int^ \infty_\infty p_1(x)x^2dx + b\int^ \infty_\infty p_2(x)x^2dx  = a\sigma^2_1 + b\sigma^2_2
\end{equation}

As the mean of $p(x)$ equals zero, the kurtosis can be written as:

\begin{equation}
\begin{split}
    K = \frac{E[x^4]}{E[x^2]^2} = \frac{E[x^4]}{(a\sigma_1^2+b\sigma_2^2)^2} = \frac{\int^ \infty_\infty p(x)x^4dx }{(a\sigma_1^2+b\sigma_2^2)^2}  \\= \frac{a\int^ \infty_\infty p_1(x)x^4dx + b\int^ \infty_\infty p_2(x)x^4dx}{(a\sigma_1^2+b\sigma_2^2)^2} = \frac{aK_1\sigma_1^4 + bK_2\sigma_2^4}{(a\sigma_1^2+b\sigma_2^2)^2}
    \label{eqKurtosislong}
\end{split}
\end{equation}

where $K_1, K_2$ are the kurtosis for the two Gaussian functions, thus they equal 3. For $a=0.95$ (newly generated population) and $b=0.05$ (surviving population), the variances can be estimated to be $\sigma_1^2 = 2$ for the short-lived population and $\sigma_2^2=6$ for the long-lived population. In this case:

\begin{equation}
    K = 3*\frac{0.95 * 2^2 + 0.05*6^2}{(0.95*2 + 0.05*6)^2} = 3.47
\end{equation}

leading to a positive value of excess kurtosis.

In the main text we also introduced the possibility of phonons (heat) as the source of the large positive excess kurtosis as early times, however, numerical simulations showed that they could not be responsible for the features at long times. It is also possible to prove that the excess kurtosis at long times for two coexisting populations will not drop lower than 0 unless one of them is not Gaussian.

Starting from a two-dimensional analogue of Eq.\ref{eqKurtosislong}:

\begin{equation}
    K = \frac{E[r^4]}{E[r^2]^2} = \frac{aK_1\sigma_1^4 + bK_2\sigma_2^4}{(a\sigma_1^2+b\sigma_2^2)^2}
\end{equation}

Where $r^2 = x^2 + y^2$- Assuming these two population propagate following Gaussian diffusion, that is, $\sigma_i = 4D_it$ and $K_1 = K_2 = 2$ 

\begin{equation}
    K = 2\frac{aD_1^2 + bD_2^2}{(aD_1 + bD_2)^2}
    \label{eq7}
\end{equation}

Where a and b are the fraction of the total population $N(t)$, thus $a = \frac{N_1(t)}{N(t)}$ and $b = \frac{N_2(t)}{N(t)}$. Substituting in eq\ref{eq7}:

\begin{equation}
    K = 2\frac{(N_1D_1)^2 + (N_2D_2)^2 + N_1N_2(D_1^2 + D_2^2)}{(N_1D_1)^2 + (N_2D_2)^2 + 2N_1N_2D_1D_2}
\end{equation}

Since $D_1^2+D_2^2 > 2D_1D_2$, the numerator is larger than the denominator and $K(t)>2$ for all points in time.

\subsection{Double Gaussian analysis}

An alternative to the kurtosis analysis is an evaluation by separating the contrast arising from each population \cite{weaver_detecting_2023} and perform independent Gaussian fits to each population. However, reliably isolating two or more overlapping populations is often challenging. In our system at sufficiently high fluences, this separation becomes feasible via a simultaneous double Gaussian fit

The results of this analysis are presented in Fig. \ref{doube_gauss_SI}, alongside the discrete variable method. Figure \ref{doube_gauss_SI}a shows that the diffusivity extracted for the fast-moving population aligns closely with the value obtained from the discrete variable analysis. This agreement confirms that our discrete method is predominantly sensitive to the dynamics of the fast (exciton) population.

Furthermore, as seen in Fig\ref{doube_gauss_SI}b, the double Gaussian fit captures the characteristic transition from a diffusive to a subdiffusive regime for the fast population. This implies that exciton transport is slowed down, an effect we attribute to interactions with trap states. Consequently, the slower population (Fig. \ref{doube_gauss_SI}a) likely corresponds to heat diffusion. This assignment is supported by the fluence dependence: the contrast from the slow population increases with fluence, consistent with a larger thermal load. In contrast, trap-related signals would saturate at high fluences as states become filled \cite{seitz_exciton_2020}.

\setcounter{figure}{0}
\renewcommand{\figurename}{Figure}
\renewcommand{\thefigure}{S\arabic{figure}}

\begin{figure}
    \centering
    \includegraphics[width=\linewidth]{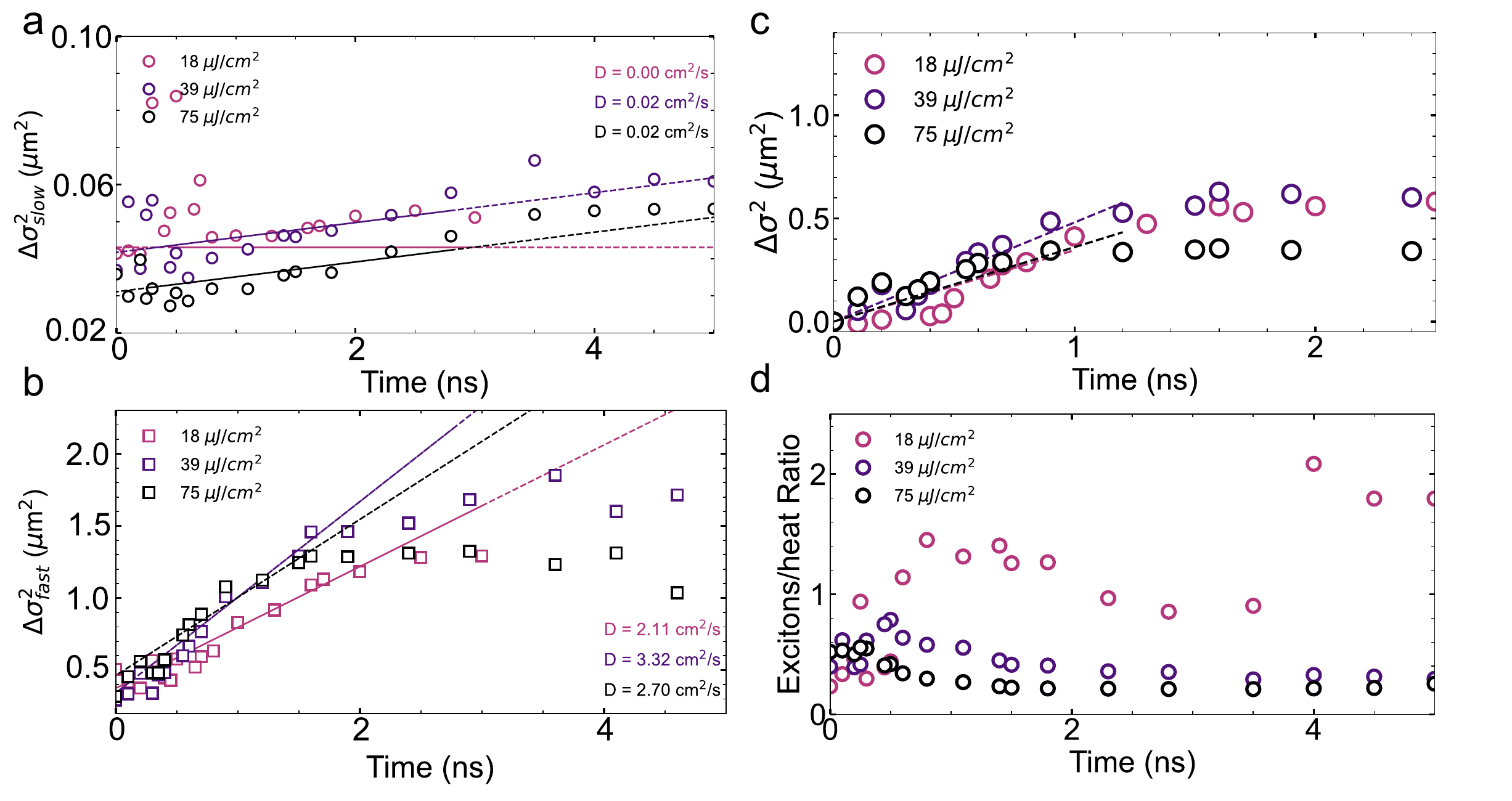}
    \caption{a. Slow component of the double Gaussian fit. b. Fast component of the Gaussian fit c. Variance extracted from the azimuthally averaged profiles using the discrete variable method d. Ratio of fast (excitons) vs slow (heat) components, the values are extracted from the double Gaussian fit}
    \label{doube_gauss_SI}
\end{figure}

\subsection{Supplementary Figures}

\begin{figure}
    \includegraphics[width=0.2\linewidth]{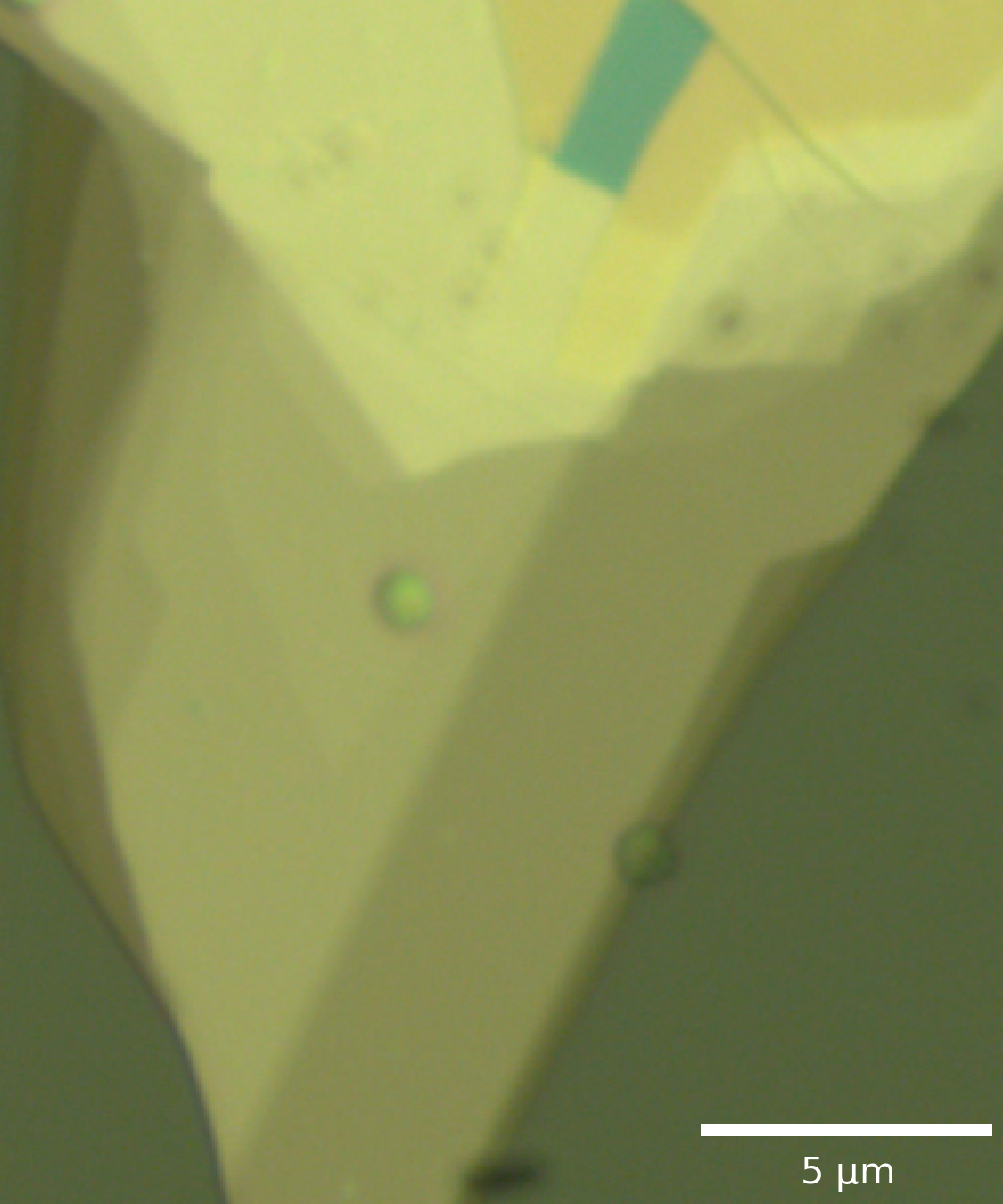}
    \caption{Brightfield image of the studied \ce{WSe2} flake.}
    \label{figS1}
\end{figure}

\begin{figure}
    \includegraphics[width=0.5\linewidth]{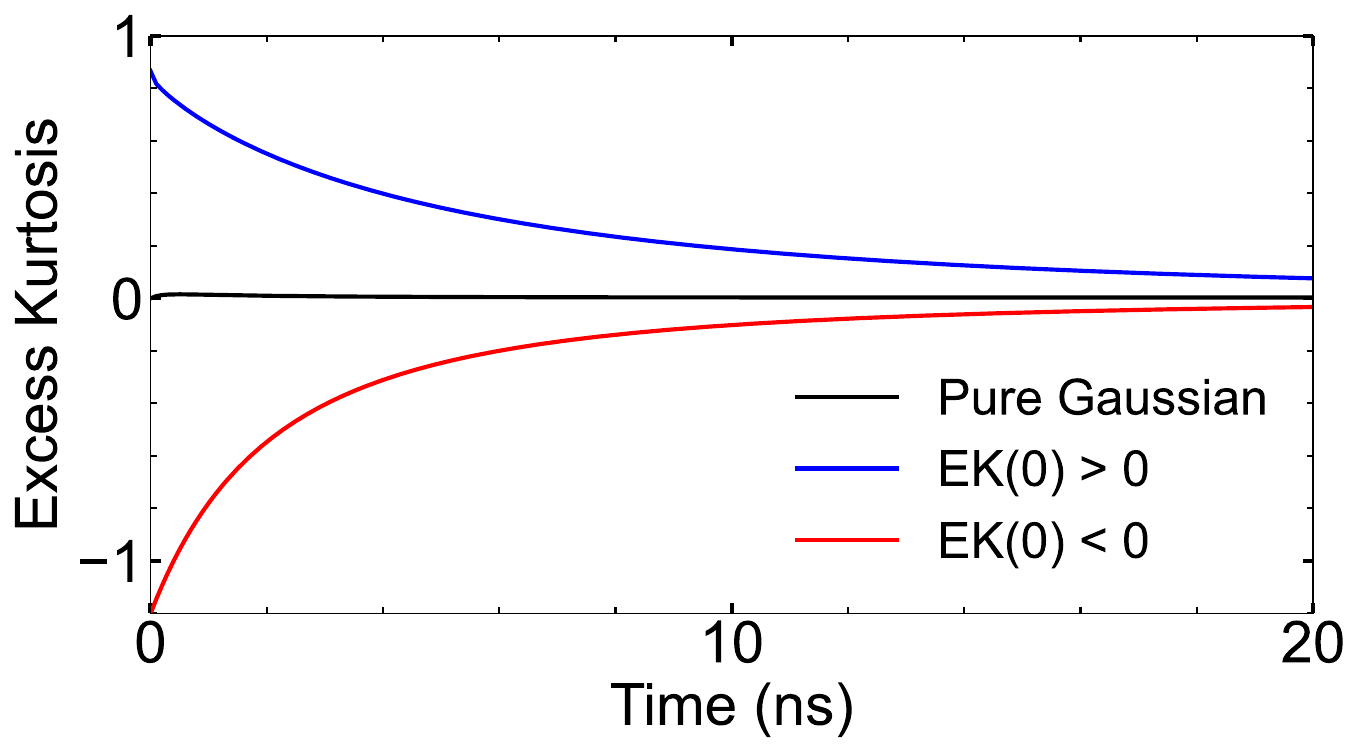}
    \caption{Evolution of the excess kurtosis for different values of initial excess kurtosis.}
    \label{figS2}
\end{figure}

\begin{figure}
    \includegraphics[width=0.5\linewidth]{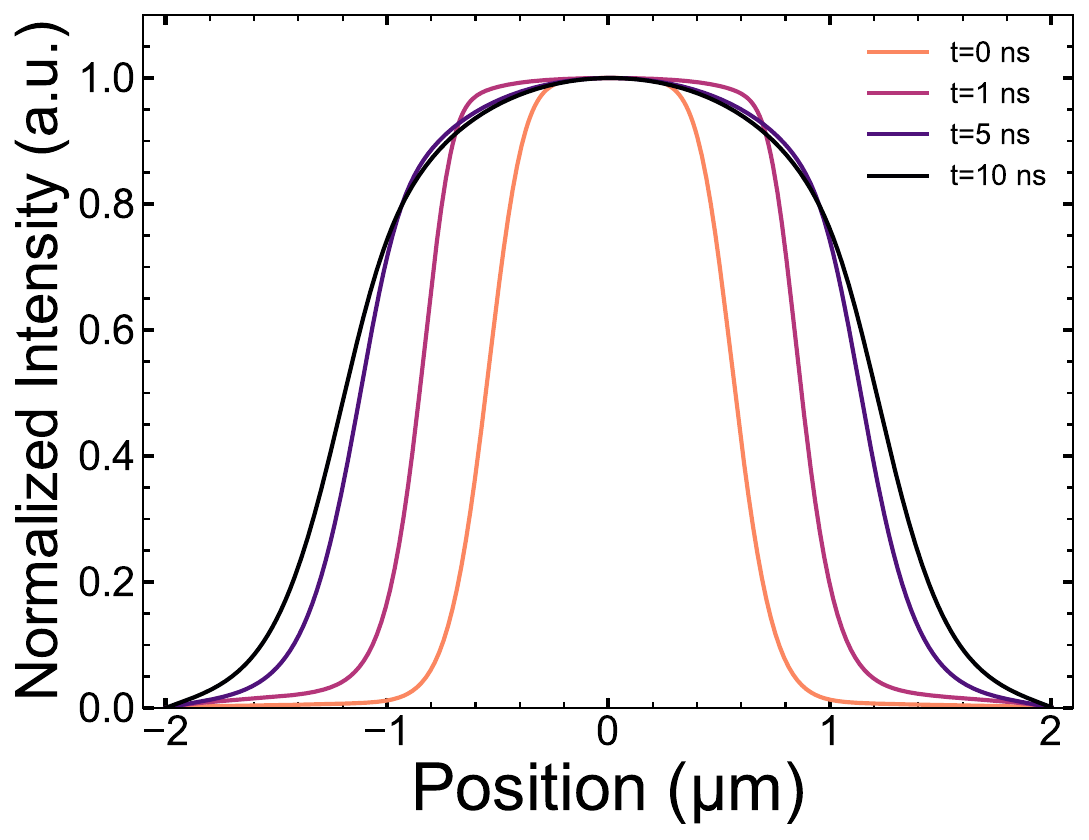}
    \caption{Numerical simulations for trap states showing the trend towards a thin-tailed distribution.}
    \label{figS3}
\end{figure}

\begin{figure}
    \includegraphics[width=0.5\linewidth]{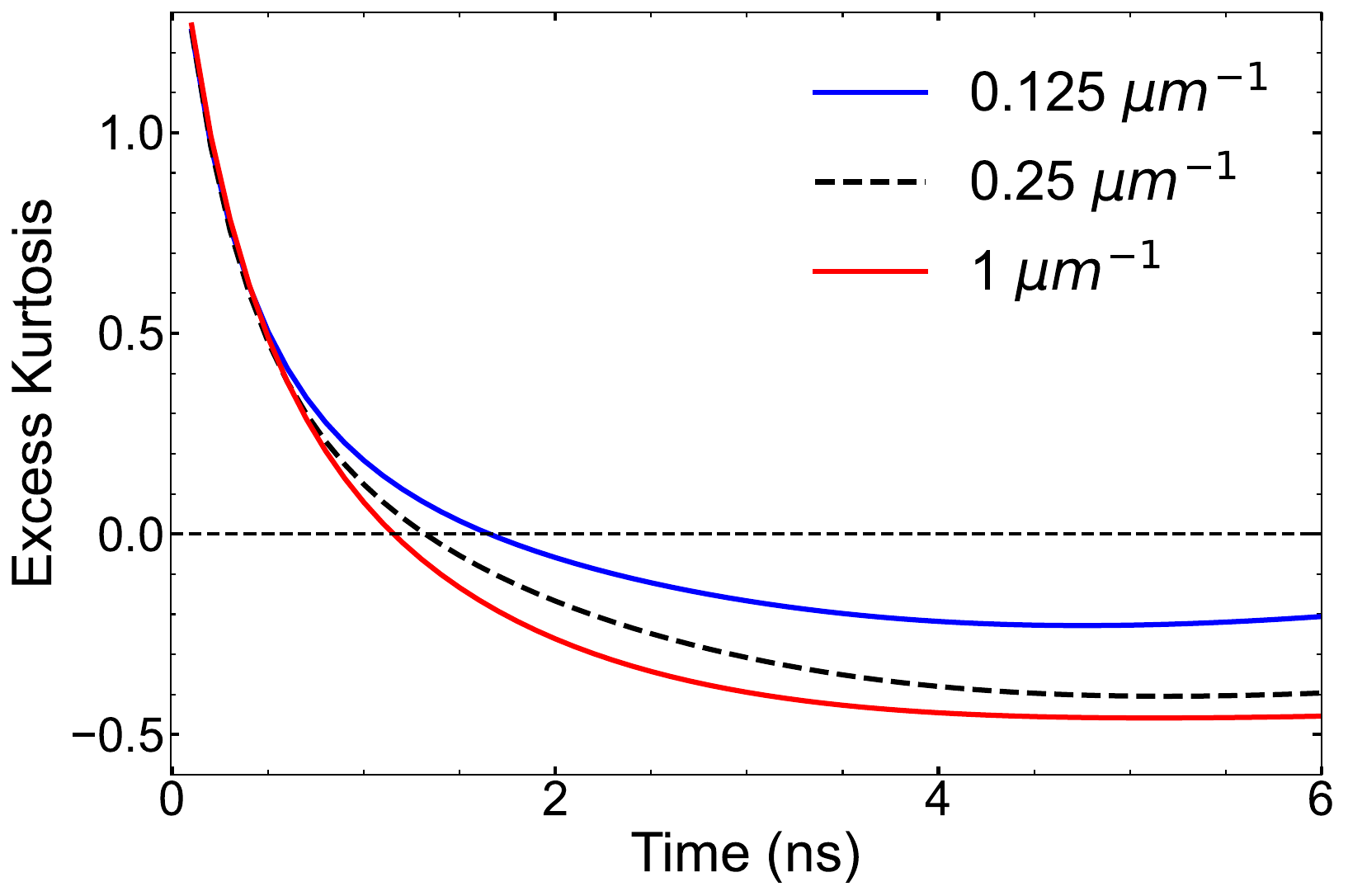}
    \caption{Numerical simulations for different values of the trap density.}
    \label{figS4}
\end{figure}

\begin{figure}
    \includegraphics[width=0.5\linewidth]{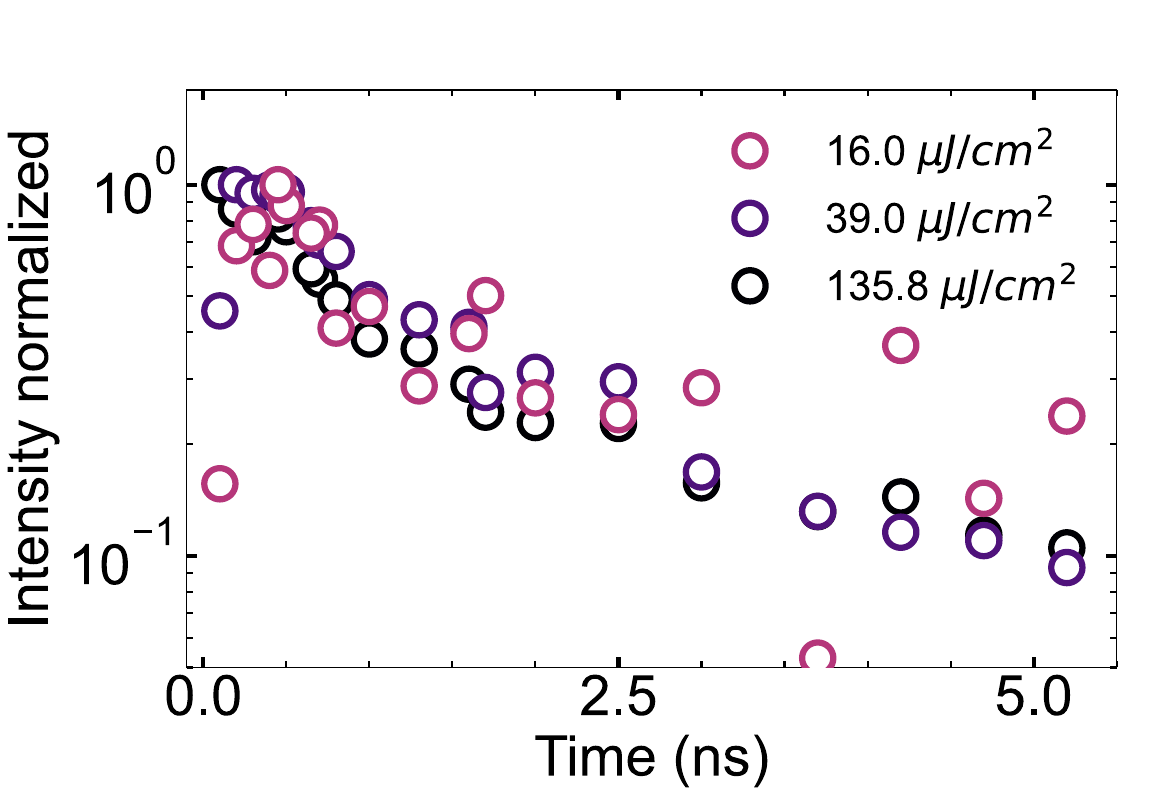}
    \caption{Lifetime traces of power-dependent data from Figure 3 in the main text.}
    \label{figS5}
\end{figure}

\begin{figure}
    \includegraphics[width=0.5\linewidth]{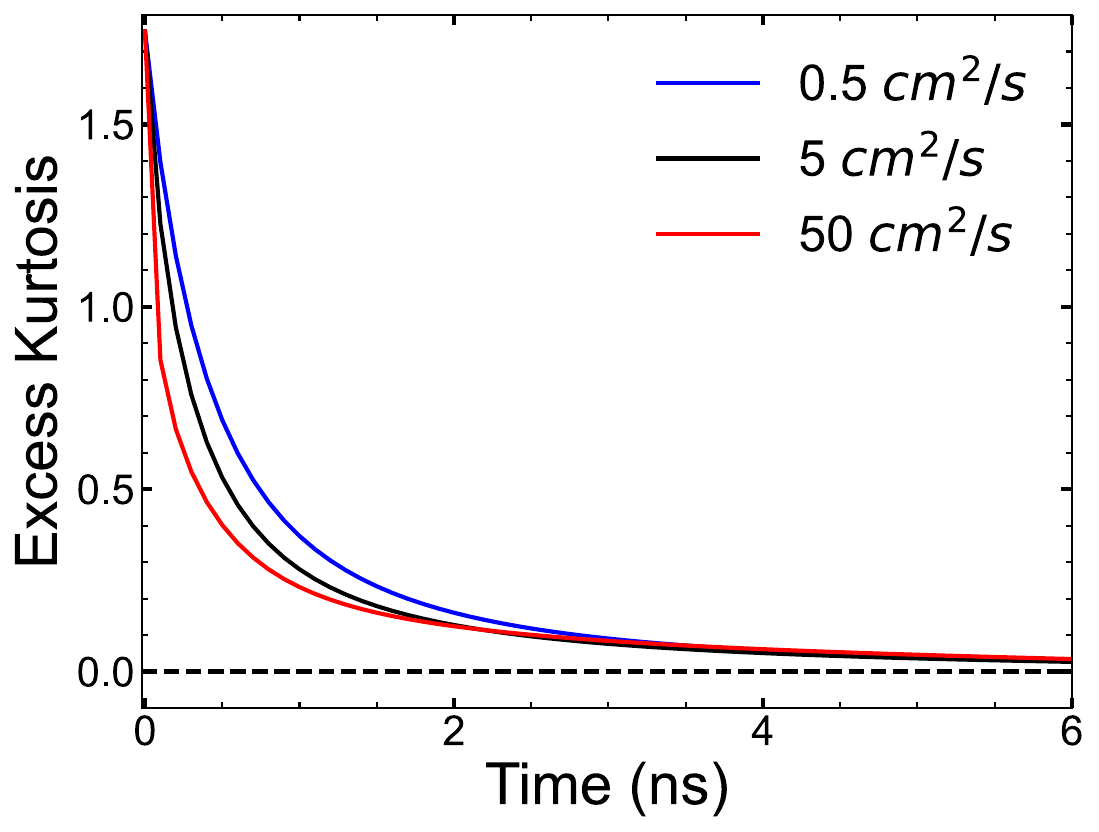}
    \caption{Evolution of the excess kurtosis for a distribution with increasing Meitner-Auger recombination rates.}
    \label{figS6}
\end{figure}

\begin{figure}
    \includegraphics[width=0.5\linewidth]{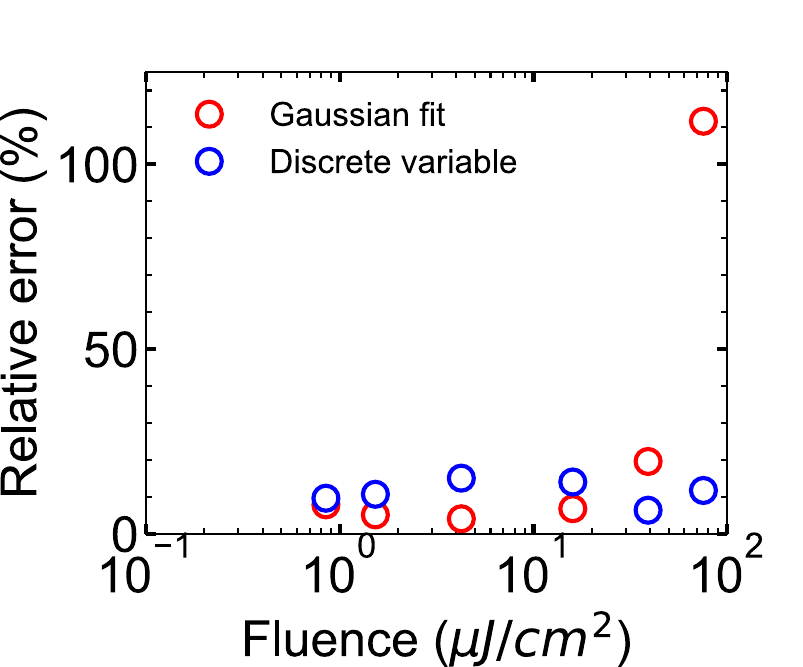}
    \caption{Relative error of the Diffusivities for both methods. The error is calculated from the fit to a linear function as depicted in the main text.}
    \label{figS7}
\end{figure}

\begin{figure}
    \centering
    \includegraphics[width=0.5\linewidth]{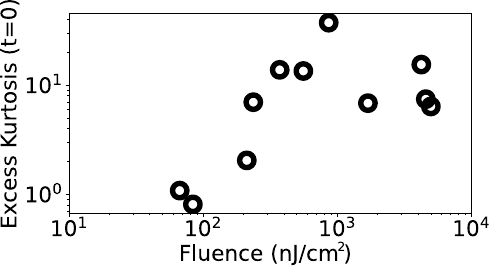}
    \caption{Excess kurtosis as a function of laser fluence. EK is calculated from the azimuthally averaged profiles to improve SNR in the low fluence measurements.}
    \label{figS8}
\end{figure}

\begin{figure}
\textbf{Table of Contents}\\
\medskip
  \includegraphics{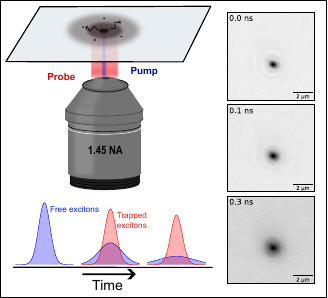}
  \medskip
  \caption{ToC Entry}
\end{figure}

\end{document}